\documentclass[journal=jctcce,manuscript=article]{achemso}
\usepackage{epsfig,graphicx}
\usepackage{amssymb,amsfonts,amsmath,bm}
\usepackage{hyperref}

\setkeys{acs}{usetitle = true}

\author{Sergei V. Krivov}
\affiliation[University of Leeds]
{Astbury Center for Structural Molecular Biology, Faculty of Biological Sciences, University of Leeds, Leeds LS2 9JT, United Kingdom}
\email{s.krivov@leeds.ac.uk}

\title{Non-Parametric Analysis of Non-Equilibrium Simulations}

\begin{document}
\maketitle

\begin{abstract}
We extend the non-parametric framework of reaction coordinate optimization to non-equilibrium ensembles of (short) trajectories. For example, we show how, starting from such an ensemble, one can obtain an equilibrium free energy profile along the committor, which can be used to determine important properties of the dynamics exactly.
New adaptive sampling approach, the transition state ensemble enrichment, is suggested, which samples the configuration space by "growing" committor segments towards each other starting from the boundary states. This framework is suggested as a general tool, alternative to the Markov state models, for a rigorous and accurate analysis of simulations of large biomolecular systems, as it has the following attractive properties. It is immune to the curse of dimensionality, it does not require system specific information, it can approximate arbitrary reaction coordinates with high accuracy and it has sensitive and rigorous criteria to test optimality and convergence.  The approaches are illustrated on a 50-dimensional model system and a realistic protein folding trajectory. 
\end{abstract}

\section{Introduction}
One general strategy in overcoming the sampling problem in biomolecular simulations consists of simulating a very large ensemble of short trajectories rather than a singe long trajectory. This strategy allows seamless parallelization and is a promising approach towards simulations employing exascale or cloud computing \cite{kohlhoff_cloud_2014, lohr_abeta_2021}. Adaptive sampling approaches can be considered as an extension of this strategy, where one, for example, improves sampling in less sampled parts of configuration space, or parts that produce largest error or controls the exploration/exploitation balance \cite{krivov_potential_2002,singhal_error_2005,doerr_learning_2014,wan_adaptive_2020,perez_adpative_2020,betz_how_2019,thomas_msm_2020}. The swarms of trajectories \cite{pan_string_2008, lev_string_2017} is another successful variation of this idea.

To analyze such ensembles of short trajectories one commonly employs the Markov state model (MSM) framework \cite{chodera_markov_2014, wu_vamp_2017, jung_artificial_2019, wan_adaptive_2020}. Assuming that the sampling is sufficiently extensive, and using a fine-grained clustering of the configuration space of the system, one can estimate the transition probability matrix. For large system, as configuration space size grows exponentially with system size, dimensionality reduction is performed before clustering. Knowing the matrix, one can compute many important properties of the equilibrium dynamics, for example, the equilibrium probabilities/populations, fluxes and rates. The minimal lag time when a MSM becomes approximately Markovian, which can be estimated by the convergence of implied timescales or by Chapman-Kolmogorov criterion, is a good indicator of the accuracy of the constructed model. The shorter is the lag time, the shorter are the trajectories, required to construct the MSM, and the larger is the possible speedup over a direct, brute-force simulation. State of the art approaches have lag times in the range of tens of nanoseconds  \cite{schwantes_improvements_2013, perez_identification_2013, hernandez_variational_2018, mardt_vampnets_2020}.

Recently, we have suggested non-parametric approaches \cite{banushkina_nonparametric_2015, krivov_protein_2018, krivov_blind_2020}, which can determine the committor and a few slowest eigenvectors, that pass stringent validation tests at much shorter lag time of trajectory sampling interval of $0.2$ ns. The non-parametric approaches are blind \cite{krivov_blind_2020} as they use no system specific information and thus do not require an extensive expertise with the system. In particular, they do not require a functional form with many parameters to closely approximate a reaction coordinate (RC), e.g., a linear combinations of molecular descriptors or a deep neural network, and can approximate any RC with high accuracy. The approaches however were restricted to long equilibrium trajectories. Here we report the extension of the framework to non-equilibrium ensembles of trajectories and suggest it as a general framework, alternative to the MSM, for a rigorous and accurate analysis of dynamics of large biomolecular systems.

We describe approaches which can be used to analyze such non-equilibrium ensembles of trajectories, to determine the following important descriptors/properties of the dynamics: the committor function, the re-weighting factors (related to the equilibrium probabilities), the eigenvectors of the equilibrium and non-equilibrium transfer operators and that of the transition probability. In particular, we show how one can determine the equilibrium free energy profile as a function of the committor, which can be used to determine exactly such important properties of the dynamics as the equilibrium flux, the mean first passage times, and the mean transition path times between any two points on the committor \cite{krivov_protein_2018, banushkina_optimal_2016}. 

One way to analyze non-equilibrium simulations consist in computing the re-weighting factors first, and then use them to re-weight the sampling, thus essentially reducing the problem to the equilibrium case \cite{wu_vamp_2017}. Such a straightforward approach, however, has the following shortcoming. The accuracy of the analysis depends on the accuracy of the obtained re-weighting factors. Thus, it requires an approach capable of determining the re-weighting factors for every trajectory point robustly and accurately, which is a very difficult task. Here, we present approaches that do not assume the existence of the re-weighting factors, and thus free of the shortcoming.

The paper is as follows. We start by reviewing the non-parametric framework for equilibrium simulations. Next, equations to determine the committor function from non-equilibrium simulations are derived. They are followed by derivation of equations to determine re-weighting factors. The power of the developed approaches is illustrated on two examples: a 50-dimensional model system and a realistic protein folding trajectory. Next, we discuss a number of realistic practical scenarios of how the developed approaches can be combined with existing enhanced/adaptive sampling techniques. We then describe a generic adaptive sampling approach, the transition state ensemble enrichment, TSEE, which is based on the developed non-parametric approaches. The performance of the TSEE is illustrated on the 50-dimensional model system. We end with a concluding discussion.

\section{Method}
A rigorous way to analyze dynamics produced by biomolecular simulations is to describe it as a diffusion on a free energy landscape, free energy as a function of RCs. The simulation trajectory is projected onto a RC by computing the RC time-series $r(t)$ as a function of time, which is used to determine the corresponding free energy landscape and diffusion coefficient. For such a description to be quantitatively accurate, the RCs should be chosen in an optimal way \cite{krivov_diffusive_2008, krivov_is_2010, krivov_protein_2018}. The committor function is an example of such a RC, that can be used to compute some important properties of the dynamics exactly \cite{krivov_reaction_2013,krivov_protein_2018}. The eigenvectors (EVs) of the transfer operator are another example \cite{mcgibbon_identification_2017}. Recently we have developed non-parametric approaches to accurately determine such coordinates from a long equilibrium trajectory \cite{banushkina_nonparametric_2015, krivov_protein_2018, krivov_blind_2020}. Here we describe how to extend this framework to non-equilibrium simulations, making possible to use these approaches for enhanced/adaptive sampling. 

\subsection{Iterative non-parametric optimization of reaction coordinates}
The overall idea of iterative non-parametric RC optimization is as follows \cite{banushkina_nonparametric_2015, krivov_protein_2018, krivov_blind_2020}. We start with a seed RC time-series $r(t)$. During each iteration we consider a variation of RC as $r(t)+\delta r(t)$, where $\delta r(t)$ can be (time-series of) any function of configuration space, collective variables and the RC itself. For example, one can take $\delta r(t)=f(r(t),y(t))$, where $y(t)$ is time-series of a randomly chosen coordinate of configuration space $X_i(t)$ or a randomly chosen collective variable and $f(r,y)=\sum_{lm} \alpha_{lm}r^ly^m$ is a low degree polynomial. The coefficients/parameters of the variation are chosen such that $r(t)+\delta r(t)$ provides the best approximation to the target optimal RC (e.g., committor). Specifically, they deliver optimum to a specific target functional $\alpha^\star=\arg \min_\alpha I(r+\delta r)$. The RC time-series is updated $r(t)\leftarrow r(t)+\delta r^\star(t)$, where $\delta r^\star(t)$ is the optimal variation, i.e., $\delta r^\star(t)=\sum_{lm} \alpha^\star_{lm}r^l(t)y^m(t)$. Iterating the process one repeatedly improves the putative RC time-series by incorporating information contained in different coordinates or collective variables. Alternatively, by repeating the iterations, the target functional is optimized by considering variations along different coordinates. For the target functionals considered here, the optimal coefficients $\alpha^\star$ are found as solutions of linear systems of equations.

While each iteration may depend on the exact choice of the family of collective variables $y$ or the parametrization of the variation $\delta r(t)$, the final RC does not, since it provides optimum to a (non-parametric) target functional, when the optimization converges. In this sense such an approach is non-parametric.

If the system obeys some symmetry (e.g., the rotational and translational symmetries for biomolecules), then the optimal RC should obey the same symmetry. A simple way to ensure this is to use as $y$, variables that respect the symmetry, for example, the  distances between randomly chosen pairs of atoms $y(t)=r_{ij}(t)$. 

The equations for non-parametric RC optimization are derived in the following sequence of steps. We first find a variational principle when the system dynamics is described by a finite Markov chain (a MSM). Next the variational principle is reformulated in terms of RC time-series. If this is possible, it means, that there is no need to construct/consider a finite Markov chain and one can operate using just RC time-series. By varying the functional we obtain the final equations for the optimal values of the parameters $\alpha^\star$. 

We employ the framework of finite Markov chains to describe Markov dynamics in the configuration space. While the configuration space in molecular simulations is continuous and rigorous treatment requires the usage of integral operators, we prefer finite Markov chains due to their convenience, simplicity and manifest invariance to the choice of coordinate systems. Moreover, a finite Markov chain can provide an excellent approximation to continuous configuration space. Since, we consider such a Markov chain as a theoretical concept, e.g., to derive the equations, and there is no need for an actual construction of such a chain in practice, the number of states can be arbitrarily large as long as it stays finite. For example, a molecular system of interest can be embedded into a large box with boundaries at $\pm L$ along each coordinate. Each coordinate can be discretized with a very fine step of, say, $d\sim 0.01$ \AA. Since the molecular dynamics simulations are usually performed by numerical integrating the Newtons equations of motion, the dynamics, at the timescales close to the simulation time step, is Markovian in the phase space not the configuration space. At longer timescales the dynamics loses memory about the momenta and can be considered approximately Markovian in the configuration space. We assume that given simulation trajectories are recorded with such or longer sampling interval, since we are mainly interested in the determination of optimal RCs as functions of the configuration space. However, it is possible, in principle, to apply the developed approaches to determine optimal RCs as functions of phase space at a shorter sampling interval.

\subsection{NPq. Non-parametric determination of the committor from an equilibrium trajectory}
We first review the approach for equilibrium trajectories \cite{krivov_reaction_2013,banushkina_nonparametric_2015,krivov_protein_2018}. We use the following notation: $\bm{X}(i\Delta t_0)$ denotes a long equilibrium multidimensional trajectory, where $\Delta t_0$ is the trajectory sampling interval;
$x(i)$ denotes an arbitrary RC as a function of MSM state $i$, while $q(i)$ is reserved for the committor; $r(i\Delta t_0)$ is an arbitrary RC as a function of trajectory snapshot or time along the trajectory or, shortly, a function of trajectory; again $q(i\Delta t_0)$ is reserved for the committor. Here we describe how, given $\bm{X}(i\Delta t_0)$, one can determine putative time-series $r(i\Delta t_0)$, which closely approximates the committor $q(i\Delta t_0)$.

Assume that, by using a fine-grained clustering, we are able to construct an accurate Markov state model, with transition probability matrix defined as $P(i|k,\Delta t_0)=n(i|k,\Delta t_0)/n(k)$, where $P(i|k,\Delta t_0)$ is the transition probability from state $k$ to state $i$ after time-interval (lag time) $\Delta t_0$, $n(i|k,\Delta t_0)$ is the number of transition from state $k$ to state $i$ after time interval $\Delta t_0$ and $n(k)=\sum_i n(i|k,\Delta t_0)$.

The committor function satisfies the following equation
\begin{subequations}
	\label{q}
	\begin{align}
	&\sum_{i}[q(i)-q(k)]P(i|k,\Delta t_0)=0,\, \mathrm{for} \, k \ne A, B\\
	&q(A)=0;\quad q(B)=1.
	\end{align}
\end{subequations}
Consider the following optimization problem:
\begin{subequations}
	\label{dx2_msm}
	\begin{align}
	&\min_x \sum_{ij}[x(i)-x(j)]^2n(i|j,\Delta t_0)\\
	&x(A)=0;\quad x(B)=1,
	\end{align}
\end{subequations}
here, $x(i)$ is an arbitrary RC as a function of state $i$. By differentiation with respect to $x(k)$, and using the detailed balance condition $n(i|j,\Delta t_0)=n(j|i,\Delta t_0)$ one obtains Eq. \ref{q}, i.e., 
the committor function provides the minimum to the functional \ref{dx2_msm}.

Before reformulating Eq. \ref{dx2_msm} optimization problem in terms of RC time-series lets introduce a convenient abbreviation for the sums like $\sum_{i=0}^{T/\Delta t}f(i\Delta t)g(i\Delta t+\Delta t)$, where $\Delta t$ equals $\Delta t_0$ or its multiple and $T=N\Delta t_0$ is trajectory length. If $\Delta t=k\Delta t_0$, it means that only $1/k$-th fraction of points in the trajectory are used. To use all the points in the trajectory one can average over the starting point as  $1/k\sum_{j=0}^{k-1}\sum_{i=0}^{T/\Delta t}f((ik+j)\Delta t_0)g((ik+j)\Delta t_0+k\Delta t_0)$, which equals $1/k\sum_{i=0}^{T-\Delta t}f(i\Delta t_0)g(i\Delta t_0+\Delta t)$. We denote such a sum as $\sum_t f(t)g(t+\Delta t)$. Even though we will mainly use $\Delta t=\Delta t_0$, the notation allows the consideration of arbitrary lag times.

The optimization problem Eqs. \ref{dx2_msm} is translated to RC time-series $r(i\Delta t_0)$ (for lag time $\Delta t$) as
\begin{subequations}
	\label{q:eq}
	\begin{align}
	&\min_r \sum_t [r(t+\Delta t)-r(t)]^2 \label{q:eqA}\\
	&r(t)=0, X(t)\in A;\quad r(t)=1, X(t) \in B, \label{q:eqC}
	\end{align}
\end{subequations}
here $r(i\Delta t_0)$ is an arbitrary RC as a function of trajectory. The total squared displacement functional in Eq. \ref{q:eqA}, which is optimized, is referred later as $\Delta r^2$ for brevity.  Here and below we assume $\Delta t=\Delta t_0$ unless stated otherwise. The theoretical minimum value of the functional, attained for
$r=q$, equals $\Delta q^2 = 2 N_{AB}$ \cite{krivov_reaction_2013}, where $N_{AB}$ is the total number of transitions from state A to B, or from B to A. Thus, if during RC optimization $\Delta r^2/2$ reaches $N_{AB}$, it follows that the putative RC closely approximates the committor.

To satisfy the constraint Eq. \ref{q:eqC} during optimization, we, first construct the seed RC time-series that satisfies the constraint, and second, during optimization, we keep positions of these points fixed by setting $\delta r(t)=0$ for them. Lets introduce boundary indicator function $I_b(t)$, which equals 1 when point $X(t)$ belongs to a boundary and is thus fixed during optimization, and zero otherwise. $\tilde{I}_b(t)=1-I_b(t)$ is its negative. The variation of the putative time series, which keeps the positions of points/frames in boundary states fixed can be taken as 
$r(t)+\delta r(t)=r(t)+\tilde{I}_b(t)\sum_j \alpha_jf_j(t)$, where $f_j(t)$ are basis functions, that are discussed below. Optimal coefficients $\alpha^\star$, which give the best approximation to the committor for the considered variation, can be found by equating the derivative of the functional with respect to $\alpha_k$ to zero:
\begin{equation}
\sum_t[r(t+\Delta t)-r(t)+\delta r(t+\Delta t)-\delta r(t)][f_k(t+\Delta t)\tilde{I}_b(t+\Delta t)- f_k(t)\tilde{I}_b(t)]=0,
\label{q:eqg}
\end{equation}
or more compact
\begin{equation}
	\sum_t[\Delta r(t)+\Delta \delta r(t)]\Delta[ f_k(t)\tilde{I}_b(t)]=0,
	\label{q:eqg2}
\end{equation}
where operator $\Delta$ denotes the forward time difference $\Delta f(t)=f(t+\Delta t)-f(t)$. It equals the following system of linear equations
\begin{subequations}
	\label{q:eqeq}
	\begin{align}
		&\sum_j A_{kj}\alpha_j^\star=b_k\\
		&A_{kj}=\sum_t \Delta [f_k(t)\tilde{I}_b(t)]\Delta[f_j(t)\tilde{I}_b(t)]\\
		&b_k=-\sum_t \Delta r(t) \Delta [f_k(t)\tilde{I}_b(t)]
	\end{align}
\end{subequations}
As basis functions $f_j(t)$ one can take the terms of a low-degree polynomial, i.e., $r^l(t)y^m(t)$ for $l+m\le n$ and $y(t)$ is a randomly chosen coordinate of the configuration space, $y(t)=X_i(t)$, or a collective variable. To focus optimization on a particular region of RC, one can modulate the polynomial terms by a common envelop. For example, to focus optimization on points around $r_0$ on the RC, one can use $e^{-|r(t)-r_0|/d}\times r^l(t)y^m(t)$, where $d$ is some small number that defines the scale. This option is useful to optimize regions corresponding to free energy minima along the committor, which get exponentially shrunk \cite{krivov_protein_2018}.

Generally, the higher is the degree of the polynomial, the faster is the optimization, though more computationally demanding. However a very high degree may lead to numerical instabilities and strong overfitting. The following strategy was found useful: use a polynomial $f(r,y)$ with a relatively small degree (3-6) for updates involving $r(t)$ and $y(t)$ followed by a polynomial $f(r)$ of a high degree (e.g., 16) for updates involving only $r(t)$.

The basic algorithm (which we call NPq) is as follows. \textbf{Initialization:} a seed RC is constructed, which satisfies the boundary constrains, for example, $r(t)=0$ if $X(t)\in A$, $r(t)=1$ if $X(t)\in B$ and $r(t)=0.5$ otherwise.  \textbf{Iterations:} one selects times-series $y(t)$ (a randomly chosen coordinate of configuration space $X$ or a collective variable), computes basis functions, solves Eqs. \ref{q:eqeq} and updates $r(t)$. \textbf{Stopping:} iterations stop when $\Delta r^2/2$ is close to the target value of $N_{AB}$ - the number of transitions from state $A$ to state $B$ or that from $B$ to $A$.

If the system has been extensively sampled, and overfitting is not possible, then, as putative $\Delta r^2/2$ reaches $N_{AB}$, the RC should closely approximate the committor. To confirm that, one can use the the $Z_{C,1}$ validation/optimality criterion for the committor \cite{krivov_reaction_2013}. $Z_{C,1}$ can be straightforwardly computed from time-series $r(i\Delta t_0)$:  each transition of trajectory from $x_1=r(i\Delta t)$ to $x_2=r(i\Delta t+\Delta t)$ adds $1/2 |x_1-x_2|$ to $Z_{C,1}(x,\Delta t)$ for all points $x$ between $x_1$ and $x_2$ \cite{krivov_reaction_2013, krivov_protein_2018}.  \textbf{Validation:} If a putative RC closely approximates the committor, then $Z_{C,1}(x,\Delta t)\approx N_{AB}$ for all $x$ and $\Delta t$, where $Z_{C,1}(x,\Delta t)$  are computed using transition path segment summation scheme \cite{krivov_reaction_2013}. \textbf{Optimality:} for a suboptimal RC, $Z_{C,1}(x,\Delta t)$ values generally decrease to the limiting value of $N_{AB}$, as $\Delta t$ increases. The larger the difference between $Z_{C,1}(x,\Delta t_1)$ and $Z_{C,1}(x,\Delta t_2)$ the less optimal the RC around $x$. Jupyter notebooks illustrating usage of $Z_{C,\alpha}$ profiles for RC analyses and, in particular, as the committor and eigenvector criteria are available at  \href{https://github.com/krivovsv/CFEPs}{https://github.com/krivovsv/CFEPs} \cite{CFEP}.

For realistic systems with limited sampling, this simple algorithm may start to overfit the RC in some regions and underfit in other. One way to overcome this problem is to make optimization adaptive, by focusing optimization on less optimized spatio-temporal regions \cite{krivov_protein_2018}. Here we consider another strategy - to use adaptive sampling in order to improve sampling in regions that are overfit or undersampled. Since adaptive sampling is no-longer equilibrium, and the described approach assumes the detailed balance, we describe a new approach applicable to non-equilibrium sampling. 

\subsection{NPNEq. Non-parametric determination of the committor from non-equilibrium sampling}

We assume that we are given a non-equilibrium ensemble of (short) trajectories. While each trajectory was simulated by following the unperturbed or natural dynamics of interest, the starting configurations are chosen arbitrarily, for example, according to an enhanced or adaptive sampling scheme. 

We employ the following representation of a non-equilibrium ensemble of trajectories. All the short trajectories are concatenated into a single long trajectory $\bm{X}(i\Delta t_0)$ combined with $\mathrm{itraj}(i \Delta t_0)$ index function which maps frames to the trajectory numbers they belong to. Our aim is to determine putative time-series $r(i\Delta t_0)$, which closely approximates the committor $q(i\Delta t_0)$.

Analysis of such non-equilibrium ensembles of trajectories by the MSM formalism is carried out without modification. One determines the transition numbers $n(i|j,\Delta t_0)$ and transition probability matrix $P(i|j, \Delta t_0)$, which can be used, e.g., to determine the committor using Eq. \ref{q} or the equilibrium probability. The non-parametric approach, however, needs modifications, as the detailed balance is not satisfied in such non-equilibrium ensembles, i.e., $n(i|j,\Delta t_0)\ne n(j|i,\Delta t_0)$ and the minimum of Eq. \ref{dx2_msm} is no longer provided by the committor function. To find a functional for non-equilibrium case, i.e., a functional whose minimum is provided by the committor function, which does not assume the detailed balance and which can be expressed in terms of RC time-series, we used the following trick. Consider the following optimization problem,
\begin{subequations}
	\label{q:imsm}
	\begin{align}
	&\left.\min_x \right|_{x'=x} \sum_{ij}[x'(i)-x(j)]^2n(i|j,\Delta t_0)\\
	&x(A)=0,\quad x(B)=1,
	\end{align}
\end{subequations}
where $\displaystyle \left.\min_x \right|_{x'=x}$ sign means that we optimize by varying $x$, while  variables $x'$ are fixed during optimization and are updated as $x'=x$ straight after, then the optimization cycle is repeated until converged. For example, assume that we minimize the functional by the steepest-descent algorithm (SD), i.e., by iteratively making steps against the gradient: $x(k)=x(k)-\mathrm{\gamma} \nabla_k$, where $\nabla_k$ is gradient and $\gamma$ is the step size. The SD will stop when the gradient is zero
\begin{eqnarray}
	\label{nabla}
&\nabla_k=-2\sum_i [x'(i)-x(k)]n(i|k,\Delta t_0)=0, \, \mathrm{for} \, k \ne A, B
\end{eqnarray}
Now, introduce the update of $x'$ variables, after every SD step, as $x'=x$. Since, we iteratively decrease a positive functional, the process should converge, hence we let $x'=x$ in Eq. \ref{nabla}, and obtain that $x$ is the committor (Eq. \ref{q}). 

Before translating Eq. \ref{q:imsm} functional to RC time-series terms, we update our notation to take into account summation over trajectories in the ensemble. Consider sum $\sum_{k=1}^{N_\mathrm{tr}}\sum_{i=0}^{T_k-\Delta t}f_k(i\Delta t_0)g_k(i\Delta t_0+\Delta t)$, where the first sum with index $k$, is the sum over $N_\mathrm{tr}$ trajectories in the ensemble and the second sum with index $i$, is the sum along $k$-th trajectory with length $T_k$. We denote such a sum as
$\sum_t f(t)g(t+\Delta t)I_t(t)$, where the sum over $t$, is the sum over the long trajectory obtained by concatenating all the trajectories in the ensemble, and $I_t(t)$ is indicator function, which equals 1 when $f(t)$ and $g(t+\Delta t)$ belong to the same short trajectory, i.e., $\mathrm{itraj}(t)=\mathrm{itraj}(t+\Delta t)$, and zero otherwise. $I_t(t)$ kills all the cross-trajectories terms, ensuring that only terms, where $f(t)$ and $g(t+\Delta t)$ are from the same trajectory, contribute to the sum. This short, intuitive notation, makes equations below less cluttered.

The optimization problem of Eq. \ref{q:imsm} is translated to RC time-series terms as follows
\begin{subequations}
	\label{q:i}
	\begin{align}
	&\left.\min_r \right|_{r'=r}\sum_t[r'(t+\Delta)-r(t)]^2I_t(t)\\
	&r(A)=0,\quad r(B)=1
	\end{align}
\end{subequations}
 Taking RC variation as $r(t)+\delta r(t)=r(t)+\tilde{I}_b(t)\sum_j \alpha_jf_j(t)$ one obtains for the gradient
\begin{equation}
	 \partial/\partial\alpha_k=-2\sum_t[r'(t+\Delta)-r(t)-\delta r(t)]f_k(t)\tilde{I}_b(t)I_t(t)
	 \label{npneqr1}
\end{equation}
Instead of optimizing with the SD, which converges rather slow, one can find analytically $\alpha^\star$, the optimal values of $\alpha$, where $\partial/\partial \alpha_k=0$:
\begin{subequations}
	\label{q:a1}
	\begin{align}
	&\sum_j A_{kj}\alpha_j^\star=b_k\\
	&A_{kj}=\sum_t f_k(t)f_j(t)\tilde{I}_b(t)I_t(t)\\
	&b_k=\sum_t[r'(t+\Delta t)-r(t)]f_k(t)\tilde{I}_b(t)I_t(t)
	\end{align}
\end{subequations}
One may attempt to speed up the convergence of the iterations further by letting $r'=r+\delta r$ in Eq. \ref{npneqr1}. Which leads to the following system of linear equations
\begin{subequations}
	\label{q:a2}
	\begin{align}
	&\sum_j A_{kj}\alpha_j^\star=b_k\\
	&A_{kj}=\sum_t [f_j(t)\tilde{I}_b(t)-f_j(t+\Delta t)\tilde{I}_b(t+\Delta t)]f_k(t)\tilde{I}_b(t)I_t(t)\\
	&b_k=\sum_t[r(t+\Delta t)-r(t)]f_k(t)\tilde{I}_b(t)I_t(t)
	\end{align}
\end{subequations}
In the non-equilibrium case, in contrast to the equilibrium one, the lower bound of  the $\Delta r^2$ functional is not known, because it depends on the sampling. To monitor the convergence of the optimization process here, we suggest to adopt one of the metrics in iterative equation solving - the increment size $|x_\mathrm{new}-x_\mathrm{old}|$. Since the optimization is stochastic ($y$ are selected randomly), we suggest to monitor the increment size during the last n iterations $||r-r_{-n}||=\sqrt{\sum_t [r(t)-r_{-n}(t)]^2}$, to have a representative estimate; here subscript $-n$ means $n$ iterations back. 

The basic \textbf{NPNEq} algorithm is similar to the equilibrium case and is as follows. \textbf{Initialization:} a seed RC is constructed, which satisfies the boundary constrains, for example, $r(t)=0$ if $X(t)\in A$, $r(t)=1$ if $X(t)\in B$ and $r(t)=0.5$ otherwise.  \textbf{Iterations:} one selects times-series $y(t)$ (a randomly chosen coordinate of configuration space $X$ or a collective variable), computes basis functions, solves Eqs. \ref{q:a2} and updates $r(t)$. \textbf{Stopping:} iterations stop when the change of RC time-series during the last n iterations $||r-r_{-n}||$ is sufficiently small.

We show in Appendix that Eq. \ref{q:a2} can be obtained in other ways. Using the  Galerkin condition, where one minimizes the error terms (the residuals or the deviations from 0) in Eq. \ref{q:imsm}, by making them orthogonal to the basis functions. Or, by minimizing the weighted sum of the error terms squared - a standard approach of solving system of linear equations iteratively.

\subsection{Validation criterion for the committor in the non-equilibrium case}
Here we suggest a generalization of the $Z_{C,1}$ criterion for the committor to the non-equilibrium case. Consider function $Z_q(x,\Delta t)$, whose derivative equals 
\begin{equation}
\label{q:zq}
\frac{\partial Z_q(x,\Delta t)}{\partial x}=\sum_{ij} \delta (x-x(j))[x(i)-x(j)]n(i|j, \Delta t)
\end{equation}
It can be computed from RC time-series $r(i \Delta t_0)$ as
\begin{equation}
\frac{\partial Z_q(x,\Delta t)}{\partial x}=\sum_t \delta (x-r(t))[r(t+\Delta t)-r(t)]
\end{equation}
If $x(i)$ is the committor, i.e., satisfies Eq. \ref{q}, then by summing Eq. \ref{q:zq} over $i$, one obtains that the derivative is zero for all $j$ but the boundary nodes. Which leads to the \textbf{validation criterion:} $Z_q(x,\Delta t)$ is constant for the committor function for all $x$ (but boundary nodes, see below) and $\Delta t$. Note that, in contrast to the equilibrium case, the constant value here is not informative, as it is defined by the transitions from the state A and depends on the sampling. In Appendix we show that $Z_q$ is an outgoing part of the $Z_{C,1}$ profile and $Z_{q}=Z_{C,1}$ for an equilibrium trajectory with the detailed balance.  

Note that, analogous to $Z_{C,1}$, $Z_q$ deviates from the constant value around the boundaries for $\Delta t>\Delta t_0$. The deviations can be eliminated
by employing the transition path segment summation scheme \cite{krivov_reaction_2013}.
However, since one expects the trajectories to be relatively short, the deviations are expected to be small, and we do not see a significant advantage in introducing this scheme here.

If a putative RC deviates from the committor, then $Z_q$ derivative should deviate from zero. However, it is not clear if the difference between the derivatives for two different $\Delta t$ can serve as a measure of RC sub-optimality. Here, the equilibrium $Z_{C,1}$ criterion is used for that purpose, which can obtained by re-weighting the non-equilibrium sampling, as demonstrated later. 

\subsection{NPNEw. Non-parametric determination of re-weighting factors from non-equilibrium sampling}
Another quantity of interest in non-equilibrium sampling are the equilibrium probabilities or re-weighting factors. Having determine the transition matrix one can compute the equilibrium probabilities, $\pi(i)$, as the solution of
\begin{equation}
\pi(i)=\sum_j P(i|j,\Delta t)\pi(j).
\end{equation}
Introducing re-weighting factors $w(i)$, which correct the non-equilibrium distribution $\pi(i)=n(i)w(i)$,  the equation can be written also as
\begin{equation}
w(i)n(i)=\sum_j n(i|j,\Delta t)w(j),
\label{w:i}
\end{equation}
here $n(i)=\sum_j n(j|i,\Delta t)$. For a single long equilibrium trajectory, where the number of ingoing and outgoing transitions for every node is equal, $\sum_j n(j|i,\Delta t)=\sum_j n(i|j,\Delta t)$, $w(i)=1$ is the solution - no re-weighting is necessary. 

The re-weighting factors do not represent a RC, as it makes little sense to project the dynamics on them. However, they can be determined by the developed formalism, and the terminology of the formalism will be used for consistency. In particular, we will refer to arbitrary re-weighting factors as a RC and the correct re-weighting factor as the optimal RC denoted by $w$ (analogous to $q$ for the committor). The aim here is to determine putative time-series $r(i\Delta t_0)$, which closely approximates the re-weighting factors $w(i\Delta t_0)$.

The corresponding optimization functional for Eq. \ref{w:i} is 
\begin{equation}
	\label{w:fi}
	\left.\min_x \right|_{x'=x} \sum_i n(i)x^2(i)/2 - \sum_{ij} x(i)n(i|j, \Delta t)x'(j)
\end{equation}
which is translated to RC time-series
\begin{equation}
	\label{w:rc}
	\left.\min_r \right|_{r'=r} \sum_t I_t(t)r^2(t)/2 - r(t+\Delta t)r'(t)I_t(t).
\end{equation}
Considering RC variation as $r(t)+\sum_j \alpha_j f_j(t)$ one obtains the following equations for the optimal parameters
\begin{subequations}
	\label{w:iter}
	\begin{align}
		&\sum_j A_{kj}\alpha_j^\star =b_k\\
		&A_{kj}=\sum_t [f_k(t)-f_k(t+\Delta t)]f_j(t)I_t(t)\\
		&b_k=-\sum_t [f_k(t)-f_k(t+\Delta t)]r(t)I_t(t)\\
		&A_{1j}=\sum_t f_j(t)I_t(t)\\
		&b_1= \sum_t 1 -r(t)I_t(t)
	\end{align}
\end{subequations}
The re-weighting factors are defined up to an overall factor, which we fix by requiring the total weight to be equal that of an equilibrium trajectory, i.e., $\sum_t w(t)I_t(t)=\sum_t 1$. This leads to Eqs. \ref{w:iter}d-e. They should replace equations Eqs. \ref{w:iter}b-c for $k=1$, for constant basis function $f_1(t)=1$, for which Eqs. \ref{w:iter}b-c give zeros. 

The re-weighting factors can also be considered as the first right eigenvector (with eigenvalue $\lambda$=1) of a non-equilibrium version of the transfer operator $n(i|j,\Delta t)/n(i)$. Appendix discusses the corresponding equations for the eigenvectors of the transfer operator $P(i|j, \Delta t)\pi(j)/\pi(i)$ and the transition probability $P(i|j, \Delta t)$. 

The optimization of eigenvectors, and, correspondingly, of the re-weighting factors, has an inherent instability \cite{krivov_blind_2020}. For example, if time-series $y(t)$, which is used to improve putative re-weighting factors, enters a region in configuration space, but does not come back, it will try to increase the weight of this region infinitely. Short trajectories are likely to make this situation more probable. To make the optimization of the re-weighting factors robust, one may need to employ some ideas discussed in \cite{krivov_blind_2020} and this is a work in progress. Here we suggest to use a selected set of proper collective variables that sample all the regions extensively, i.e., they contain transitions to and from all the sampled regions. 

In the simplest case one can take as $y(t)$ only a single committor coordinate time-series. In this case one will determine $w(q)$, re-weighting factors as a function of the committor. This is sufficient, for example, for the first passage ensemble which consists of trajectories starting in A and stopping as soon as they reach B, since the biasing factor in this ensemble is a function of the committor. It should also be sufficient for an ensemble of short trajectories for a system with a single dominant pathway. In this case, the committor function, increasing along the pathway, can be used to parameterize the pathway and the re-weighting factors. If there are two (or a few more) parallel pathways one can incorporate a proper collective variable that distinguishes between them into optimization as $y(t)$.

The basic \textbf{NPNEw} algorithm is as follows. \textbf{Initialization:} a seed RC is initialized to $r(t)=1$. \textbf{Iterations:} one randomly selects  times-series $y(t)$ from a set of proper collective variables, computes basis functions, solves Eqs. \ref{w:iter} and updates the putative RC time-series. \textbf{Stopping:} iterations stop when the change of RC time-series during the last n iterations $||r-r_{-n}||$ is sufficiently small.

Once computed, the re-weighting factors are used to determine the equilibrium properties. For example, for the equilibrium free energy profile $F(r)$: each trajectory point $r(i\Delta t_0)$ contributes with corresponding weight of $w(i\Delta t_0)$; for equilibrium $Z_{C,\alpha}$ cut-profiles: each transition from $r(i\Delta t_0)$ to $r(i\Delta t_0+\Delta t)$ contributes with corresponding weight of $w(i\Delta t_0)$.

\section{Illustrative Examples}
\subsection{50 dimensional model system}
\begin{figure}[htbp]
	\centering
	\includegraphics[width=.5\linewidth]{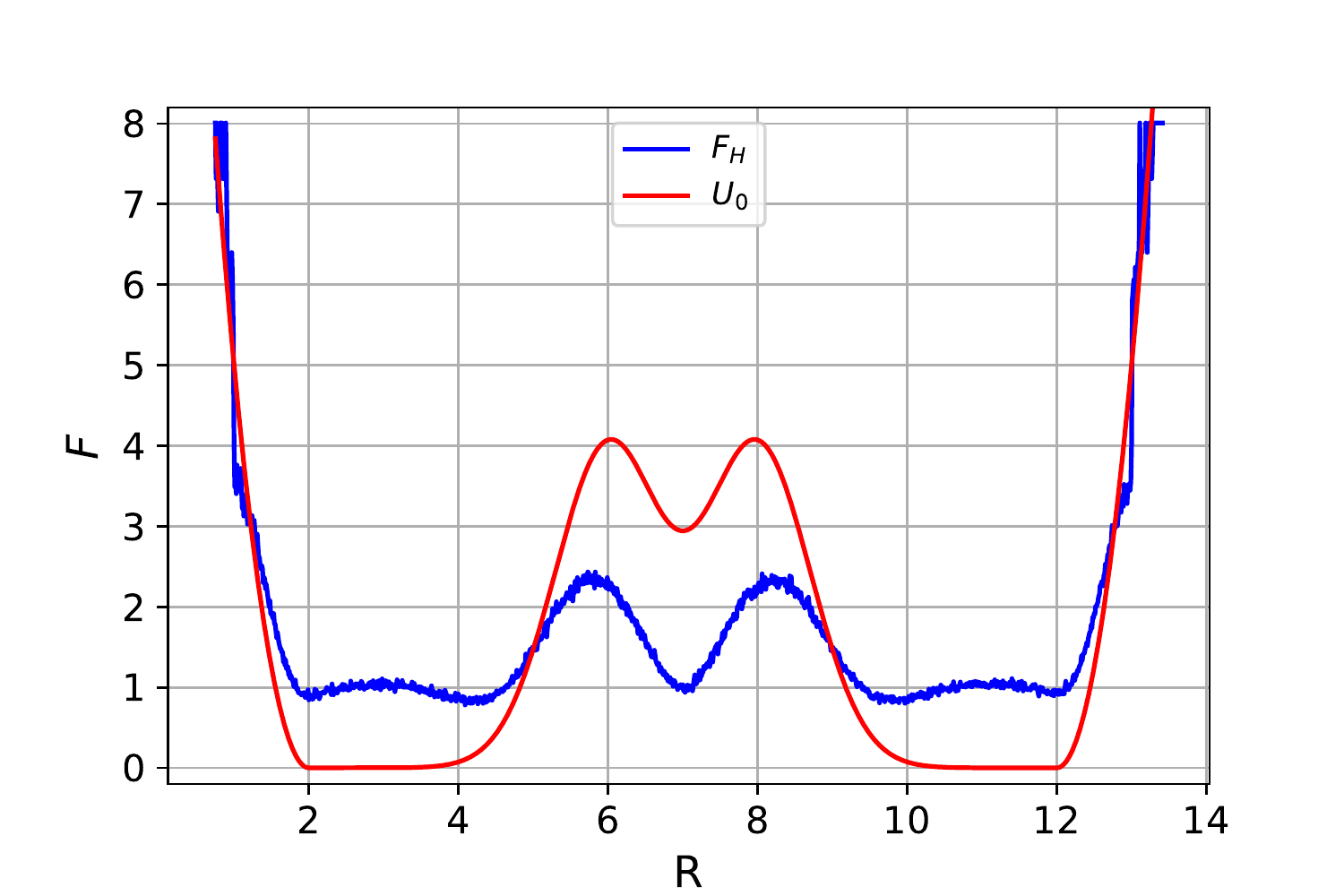}
	\caption{Non-equilibrium free energy profile $F_H(R)$ (blue) and $U_0(R)$ (red).}
	\label{fig:50FEP}
\end{figure}

As the first model system we consider a high-dimensional system for which the committor function can be computed analytically. It qualitatively resembles a protein folding landscape with radially symmetric potential energy $U(\bm{X})=U(R)$, decreasing towards the beginning of the coordinates. The decrease in enthalpy is compensated by the decrease in entropy so that the resulting free energy profile as a function of $R$ has two minima, separated by a barrier (Fig. \ref{fig:50FEP}). More specifically, $U(R)=U_0(R)-(n-1)\ln(R)$, where $R=\sqrt{\sum_{i=1}^{n} X_i^2}$ and 
\begin{equation}
	U_0(R)=
	\begin{cases}
		R<2 & 5(R-2)^2\\
		2\le R \le 12 & 4e^{-(R-6)^2} +4 e^{-(R-8)^2}\\
		12<R& 5(R-12)^2
	\end{cases}
\end{equation}

Due to high dimensionality of the configuration space, $n=50$ here, the system can not be analyzed directly by an MSM approach, one would need to preform a dimensionality reduction first. For example, a trajectory of $10^6$ frames will not even visit every possible region of configuration space with different combinations of coordinate signs. Approaches assuming pathways can not be applied also, as the system does not have a well defined pathway. 

Non-equilibrium ensemble of short trajectories was obtained by randomly selecting a point in the 50 dimensional configuration space with uniform distribution in $1<R<13$ and simulating a diffusion trajectory for 10 steps with $D(\bm{X})=1$, simulation step $\Delta t_\mathrm{sim}=0.001$ and saving interval of $\Delta t_0=0.1$. The total size of the ensemble is $10^6$ points. The free energy profile as a function of the radius $F_H(R)$, computed from the trajectories, is different from $U_0(R)$ (Fig. \ref{fig:50FEP}), confirming the non-equilibrium character of sampling.
The NPNEq algorithm is used to find the putative committor time-series. Specifically, \textbf{Initialization:} a seed RC is constructed as $r(t)=0$ if $R(t)<2$ (state A, see Fig. \ref{fig:50FEP}), $r(t)=1$ if $R(t)>12$ (state B) and $r(t)=0.5$ otherwise.  \textbf{Iterations:} Every iteration consists of four RC updates.  NPNEq equations (Eqs. \ref{q:a2}) with basis functions being the terms of polynomial $f(r,y)$ of degree 6, where $y(t)=X_i(t)$ and $i$ is randomly chosen from $1,2,...,50$, i.e., $y(t)$ is a randomly chosen coordinate time-series. It is followed by NPNEq equations with basis functions being the terms of polynomial $f(r)$ of degree 16 with envelop $\exp(-|1-r|/0.005)$, that with envelop $\exp(-|r|/0.005)$, and that without envelop. \textbf{Stopping:} iterations are terminated when $||r-r_{-100}||<0.3$. A Jupyter notebook with the analysis is provided in the Supporting Information and is also available at \href{https://github.com/krivovsv/NPNE}{https://github.com/krivovsv/NPNE} \cite{npne}.

The results are robust with respect to the polynomial degrees, frequency of updates with envelops, size of the envelops, etc. Higher degrees generally lead to faster convergence, a bit smaller value of the $\Delta r^2$ functional, and less fluctuating $Z_q$, though very high degrees may result in instability and occasional failure to converge.   

\begin{figure}[htbp]
	\centering
	\includegraphics[width=.5\linewidth]{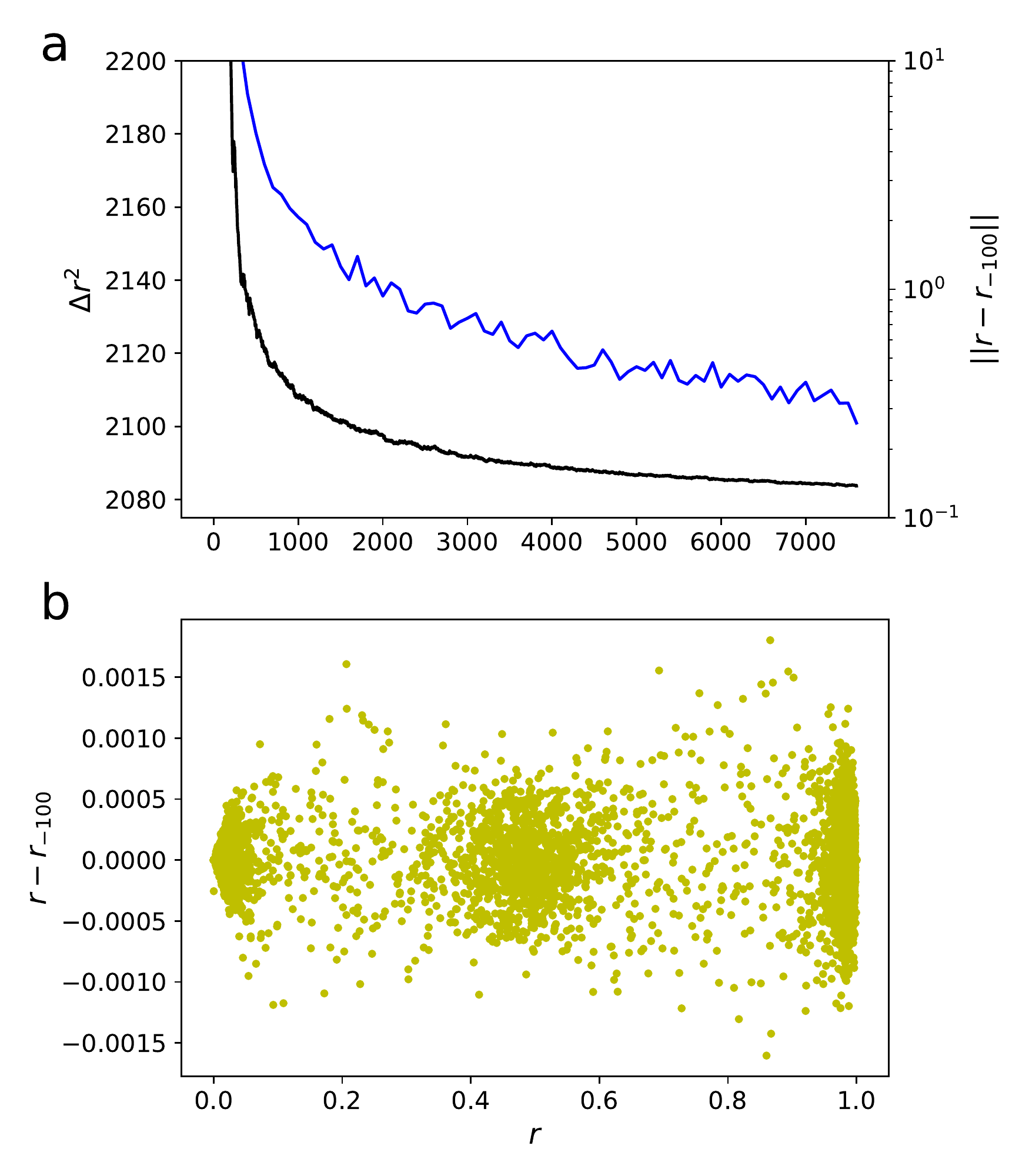}
	\caption{Convergence of the NPNEq optimization. a) $\Delta r^2$ (black) and $||r-r_{-100}||$ (blue) as functions of iteration number. b) Change of the RC time-series during the last 100 iterations for selected frames.}
	\label{fig:50conv}
\end{figure}

Fig. \ref{fig:50conv} demonstrates the convergence of the iterations of the optimization process.  The size of increments $||r-r_{-100}||$ are steadily getting smaller with the iteration number. The change in the functional $\Delta r^2$ value as a function of iteration number is getting smaller, indicates that we are approaching the minimum. The change of the RC time-series during the last 100 iterations, for selected frames, is bounded by $0.002$, indicating that the convergence is uniform.

\begin{figure}[htbp]
	\centering
	\includegraphics[width=.5\linewidth]{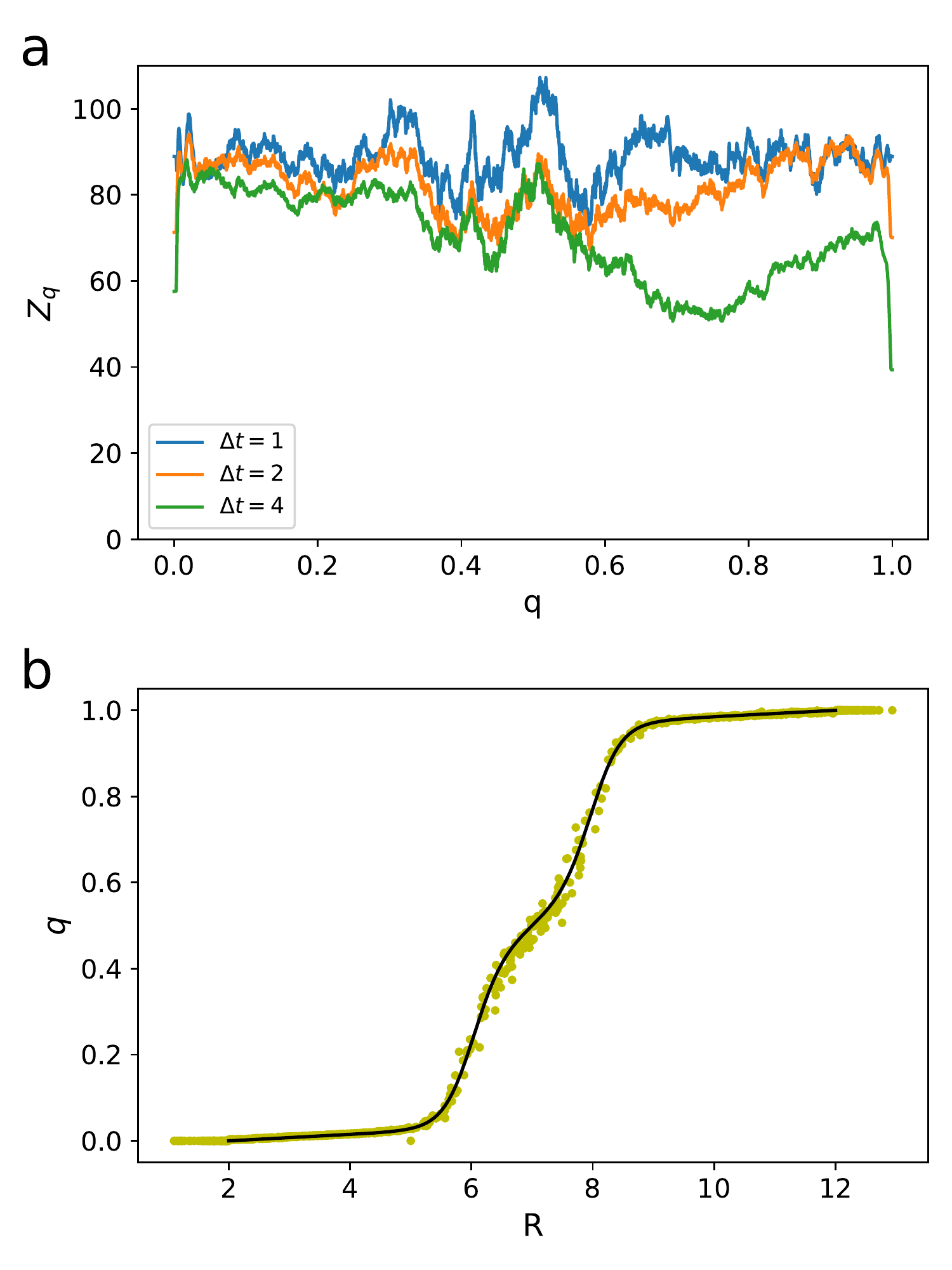}
	\caption{Validation of the putative RC time-series. a) Non-equilibrium committor validation criterion $Z_q(x,\Delta t)$ along putative time-series $q$ for $\Delta t=1,2,4$ are relatively constant. b) Comparison of the analytically computed committor as a function of $R$ (black line) with that for selected frames from RC time-series (yellow dots).}
	\label{fig:50zq}
\end{figure}

Fig. \ref{fig:50zq} inspects how closely the determined time-series approximates the committor. The validation criterion is relatively constant. The root mean squared deviations of  $Z_q$ are about 5, 6 and 10 for $\Delta t=1,2$ and $4$, respectively. Larger fluctuations for $\Delta t=4$ could be due to general statistical fluctuations because of limited sampling. Unlike the equilibrium $Z_{C,1}$ profiles, the mean values of the $Z_q$ profiles are not very meaningful, as they depend on the transitions from state A, which depend on the sampling.  The committor as a function of $R$ can be compute analytically as $$q(R)=\int_{R(A)}^R D^{-1}(x)e^{U_0(x)}dx/\int_{R(A)}^{R(B)} D^{-1}(x)e^{U_0(x)}dx,$$ where $D(x)=1$. 
Fig. \ref{fig:50zq}b shows that the latter is in a good agreement with the putative RC, which is referred as committor henceforth.

The re-weighting factors are computed using the NPNEw algorithm. Specifically, \textbf{Initialization:} a seed RC is initialized as $r(t)=1$. \textbf{Iterations:} NPNEw equations (Eqs. \ref{w:iter}) with basis functions being the terms of polynomial $f(r,y)$ of degree 5, where $y(t)$ is the putative committor time-series $q(t)$. \textbf{Stopping:} iterations are terminated when $||r-r_{-1}||<0.0001$.

\begin{figure}[htbp]
	\centering
	\includegraphics[width=.5\linewidth]{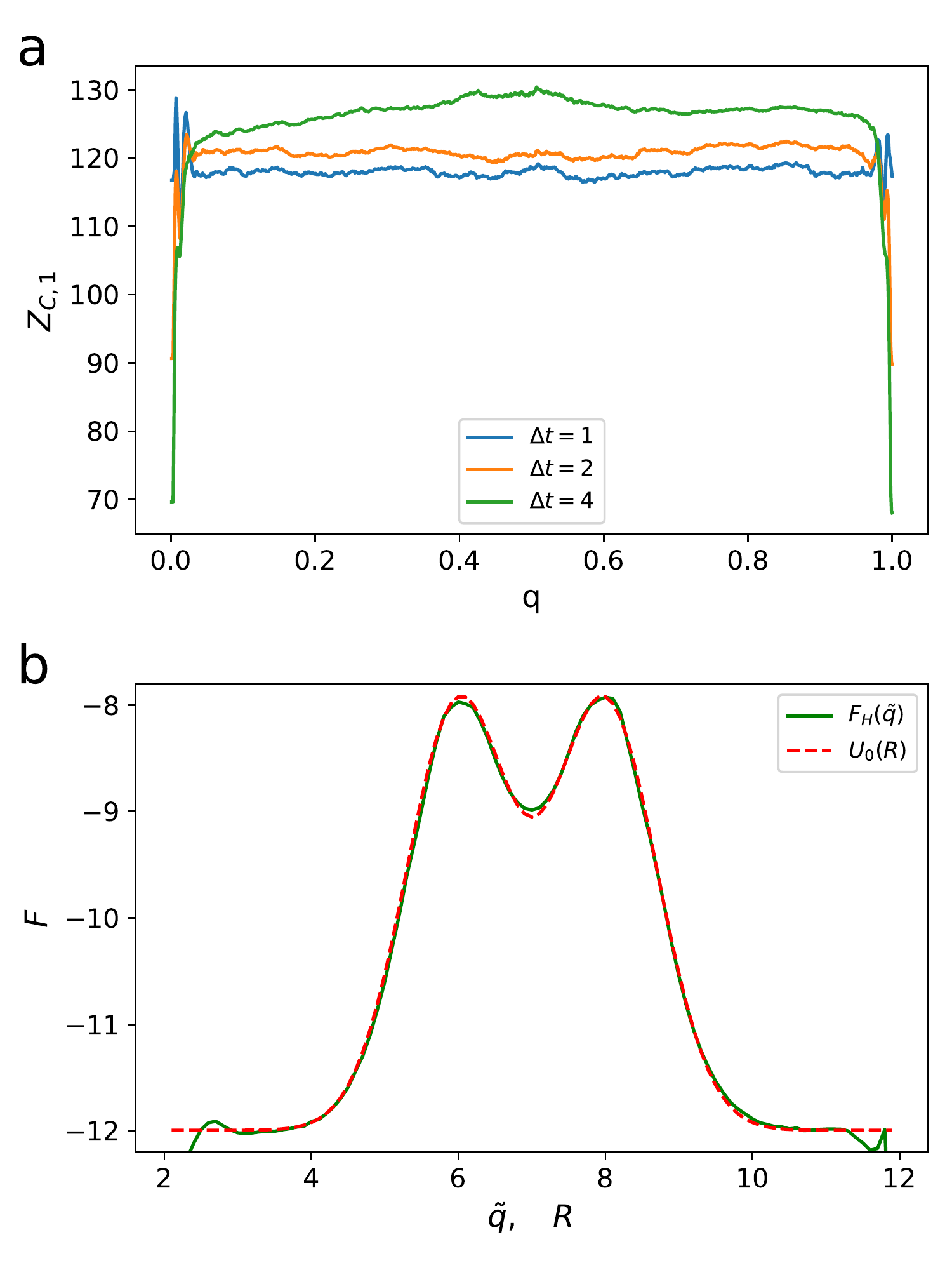}
	\caption{Equilibrium properties. a) Equilibrium validation/optimality criterion along the putative time-series is constant. The deviations from the constant value are in the range of $10\%$. b) $F_H(\tilde{q})$ (green), equilibrium free energy profile as a function of $\tilde{q}$, is in agreement with $U_0(R)$ (red); $F_H(\tilde{q})$ was shifted horizontally and $U_0(R)$ vertically for maximum overlap.}
	\label{fig:50eq}
\end{figure}

The re-weighting factors are used to compute the equilibrium properties. Fig. \ref{fig:50eq}a shows the equilibrium $Z_{C,1}$ as the function of the putative committor $q$, the committor validation criterion. The profile is constant with fluctuations bounded by $10\%$, confirming that $q$ approximates the committor rather well. 

The equilibrium $Z_{C,1}$ profile can be used to compute the equilibrium flux $J_{AB}=N_{AB}/T$, where $T=N\Delta t_0$, is the total length of trajectory and $N_{AB}$ is the number of transitions from A to B. $N_{AB}$ can be computed as $N_{AB}^{-1}=\int_{q(A)}^{q(B)} Z_{C,1}^{-1}(q)dq$ and $N=\sum_t w(t)I_t(t)$. The obtained value of the equilibrium flux $J_{AB}=0.001179$ is in a good agreement with that computed analytically as $J_{AB}=N_{AB}/Z=0.001186$, where 
$N_{AB}^{-1}=\int_{R(A)}^{R(B)} Z_{C,1}^{-1}(x)dx=\int_{R(A)}^{R(B)} D^{-1}(x)e^{U_0(x)}dx$ and $Z=\int e^{-U(x)} dx$, where $D(x)=1$.

The re-weighting factors can be used to compute the equilibrium free energy profile $F_H(q)$ and the diffusion coefficient $D(q)$ as functions of the committor and thus provide the diffusive model of the equilibrium dynamics along the committor, which can be used to compute the following important properties of the dynamics exactly: the equilibrium flux, the mean first passage times, and the mean transition path times between any two points on the committor \cite{krivov_protein_2018, banushkina_optimal_2016}. However, using $F_H(q)$ for the analysis and description of the dynamics is not very convenient as the diffusion coefficient varies significantly along the coordinate. It is more convenient to use a “natural” coordinate \cite{krivov_diffusive_2008, krivov_protein_2018}, $\tilde{q}$, where the diffusion coefficient is constant $D(\tilde{q}) = 1$. It is related to $q$ by the following monotonous transformation $d\tilde{q}/dq=D(q)^{-1/2}$. Fig. \ref{fig:50eq}b shows that $F_H(\tilde{q})$ is in a very good agreement with $U_0(R)$.

In summary, this example illustrates that the NPNEq and NPNEw algorithms can be used to determine the committor and re-weighting factors from non-equilibrium ensembles of short trajectories and to construct a diffusive model of equilibrium dynamics, which can be used to compute important properties of the equilibrium dynamics exactly.

Functions implementing NPq (Eq. \ref{q:eqeq}), NPNEq (Eq. \ref{q:a2}) and NPNEw (Eq. \ref{w:iter}) iterations, computing $Z_{q}$ and $Z_H$ profiles and performing transformation to natural coordinate are available as Python library npnelib.py at 
\href{https://github.com/krivovsv/NPNE}{https://github.com/krivovsv/NPNE} \cite{npne}.

\subsection{A realistic protein folding trajectory}
We have demonstrated that the NPNEq algorithm can accurately determine the committor RC from a non-equilibrium sampling of the model system. The model system has a relatively simple configuration space and a relatively simple committor function, which is a function of $R$ only. It is of interest to see how accurately the NPNEq algorithm can approximate the committor for a realistic system. To this end, the NPNEq algorithm is applied to a long equilibrium protein folding trajectory 
of HP35 Nle/Nle double mutant consisting of 1509392 snapshots at 380 K \cite{piana_protein_2012}, in particular, to compare with its equilibrium version \cite{krivov_protein_2018}. The analysis details can be found in a Jupyter notebook, provided in the Supporting information and at  \href{https://github.com/krivovsv/NPNE}{https://github.com/krivovsv/NPNE} \cite{npne}.

\begin{figure}[htbp]
	\centering
	\includegraphics[width=.5\linewidth]{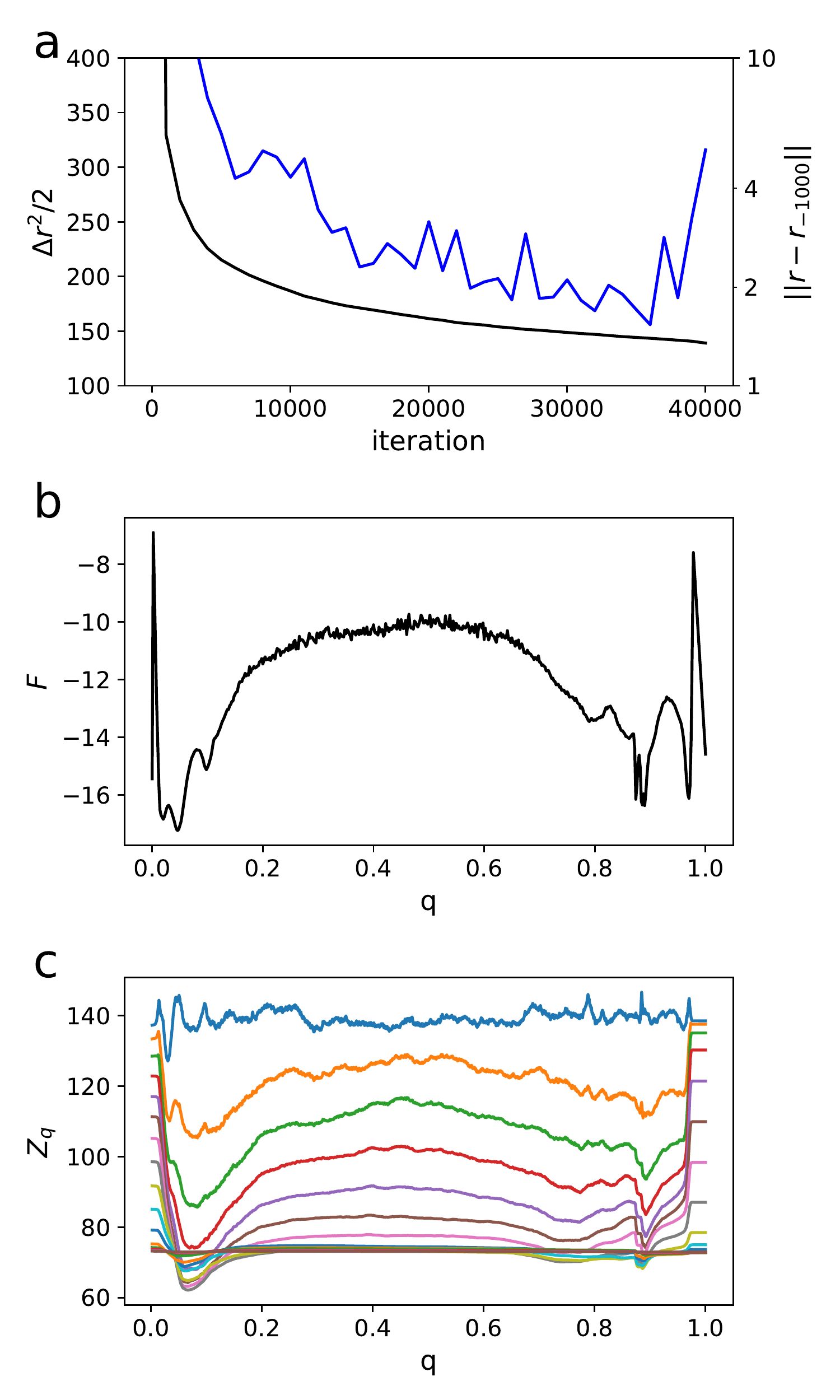}
	\caption{Application of the NPNEq to a realistic protein folding trajectory.  a)
		Convergence of the NPNEq optimization: $||r-r_{-1000}||$ (blue) and $\Delta r^2$ (black) as functions of iteration number. b) Free energy as a function of committor, $F_H(q).$ c) $Z_q(x,\Delta t)$ along putative committor time-series for $\Delta t=1$ (blue), $2$ (orange), $... , 2^{15}$.}
	\label{fig:hp35}
\end{figure}

The optimization continued for 40000 iterations. The final $\Delta r^2/2\sim 1.9N_{AB}$, i.e., almost two times higher than the target value of $N_{AB}=74.5$.
Fig. \ref{fig:hp35}a inspects the convergence of the algorithms. As once can see the increment size $||r-r_{-1000}||$ converges to some non-zero value, while the $\Delta r^2/2$ functional continues to decrease, indicating that the optimization process will overfit eventually, if continued, by going below the lower bound of $\Delta r^2/2=N_{AB}.$ 

Fig. \ref{fig:hp35}b shows that the free energy profile as the function of the putative committor, $F(q)$, is very similar to that obtained with equilibrium adaptive non-parametric optimization \cite{krivov_protein_2018}, indicating that the non-equilibrium approach has similar approximation power.

$Z_q$ criterion (Fig. \ref{fig:hp35}c), which, for equilibrium dynamics, is equivalent to $Z_{C,1}$ shows that $Z_q(q,\Delta t_0)$ is almost 2 times larger then $N_{AB}=74.5$. It means that the diffusive model of dynamics is accurate, within a factor of 2, at the time scale of $\Delta t_0=0.2$ ns. For example, the folding free energy barrier can be estimated with a error about $kT\ln(2)\sim 0.4$ kcal/mol.

The model can be improved further by continuing the optimization. However, since the sampling is limited and not extensive, it will lead to overfitting. Fig. \ref{fig:hp35}c
shows that optimization is not uniform with $Z_q(q,\Delta t_0)-Z_q(q,2\Delta t_0)$ is smallest for $0.4<q<0.6$. If continued further, $Z_q(q,\Delta t_0)$ will get lower than $Z_q(q,2\Delta t_0)$ in that region, indicating that the putative RC is overfitted around the transition state. One way to avoid overfitting, is to make optimization more uniform by focusing it on underfitted/suboptimal regions \cite{krivov_protein_2018}. An alternative approach consists in performing additional extensive sampling of the transition state by starting many short trajectories from the frames in the overfitted region and analyze the combined simulations using the developed non-equilibrium approach.

\section{Adaptive sampling}
Given a representative and extensive, possibly non-equilibrium sampling of the configuration space the proposed approaches can be used to determine the equilibrium free energy profile as a function of the committor. The later, in particular, can be used to determine important properties of the equilibrium dynamics exactly. By a representative sampling we mean such a sampling which contains all the important regions of the configuration space, e.g., all the important transition pathways, or a representative sample of them, if their number is infinite. By an extensive sampling we mean a sampling of such a size that overfitting by the non-parametric approaches is not possible or negligible. In this section we will discuss possible strategies of generating such a representative and extensive sampling.

Consider first the case where a trajectory or ensemble of trajectories provide representative, though not extensive sampling, for example, the state-of-the-art protein folding trajectories \cite{shaw_atomic-level_2010, lindorff-larsen_how_2011}. Applying the non-parametric approaches (either equilibrium or non-equilibrium) one finds that the optimization soon starts to overfit the committor RC in the TS region, because the sampling of this region is relatively poor, compare to the rest of the configuration space \cite{krivov_protein_2018, krivov_blind_2020}. The regions where the putative RC is overfitted can be detected by using the optimality criteria. In order to avoid the overfitting, many additional short simulations are performed, starting from the configurations that belong the overfitted regions, e.g., the TS region. Then, the total simulation data is analyzed by the non-parametric non-equilibrium approach.

A more difficult case is when the initial representative sampling is absent. For systems with relatively simple, small configuration space, selection of initial configurations to start many short simulations as well as the seed RC can be done analytically, as it was done for the model system considered here. Such systems may include practically important cases such as, e.g., studies of dynamics of a ligand binding/unbinding to/from a protein \cite{betz_how_2019,thomas_msm_2020}, or diffusion of a small molecule/ion through an ion channel pore.

If one of the boundary states has a much shorter lifetime compared to the other state, then many trajectories should be started from the former state, which shall generate a non-equilibrium (first passage) representative sampling. 

If both boundary states have long residence times, while the transition path times are rather short, one can use the transition path sampling approach \cite{bolhuis_transition_2002} to generate a representative sampling. Inclusion of the rejected paths will increase the size of the sampling and remove conditioning on the boundary states. 

Another possibility is to use biased, non-equilibrium sampling, though, in this case, representative sampling of transition paths is not guaranteed. For example, one may use sampling at a higher temperature or sampling with a bias potential, e.g., umbrella sampling \cite{torrie_nonphysical_1977, souaille_umbrella_2001}, steered-MD \cite{isralewitz_steered_2001}, replica-exchange \cite{sugita_replica_1999, fukunishi_replica_2002}, meta-dynamics \cite{barducci_metadynamics_2011}, or forward flux sampling \cite{allen_rare_2006, sarwar_rare_2020}. If an enhanced sampling method perturbs the dynamics of interest, e.g., a higher temperature or a biasing potential, then many short simulations with unperturbed dynamics, need to be performed, starting from the obtained configurations. 

String method using swarms of trajectories \cite{e_string_2002, pan_string_2008, lev_string_2017} can be straightforwardly combined with the NPNEq approach. Since the latter does not assume the existence of a dominant pathway, it may improve performance of the former in systems, where this assumption does not hold.

Consider now the forward flux sampling (FFS) \cite{allen_rare_2006, sarwar_rare_2020}, where one uses an order parameter (OP), which can be different from the optimal RC - the committor, to propagate the trajectories from state A to state B. While the accuracy of FFS does not depend on the OP, the efficiency does. Thus it would be desirable to propagate FFS trajectories using the optimal RC or committor. Since the committor RC is not known in advance, one possibility is to compute the committor during sampling, applying the developed approach to the data sampled so far. Having the idea in mind we propose the following approach.

We first describe an idealized scenario. Assume that relatively long unbiased simulations were performed in both boundary states A and B. The simulations are not long enough, however, for the system to sample the transitions between the states, and thus can not be used to construct the entire committor RC. Assume now that these simulations, however, can be used to construct the committor in the sampled regions, i.e., the starting and ending segments of the committor for example $[0,\alpha]$ and $[\beta,1]$. Then many short trajectories are started at
the points with committor close to $\alpha$ and $\beta$. Analyzing the combined new and old simulations, one extends the RC segments to a large value of $\alpha$ and a smaller value of $\beta$, since some of the stochastic trajectories will travel to these regions. One continues in such an iterative manner to grow the two segments towards each other, until they meet, when $\alpha=\beta$, thus providing the initial representative sampling of transition paths. 

Unfortunately, it is not possible to construct accurately just the two segments of the RC, because as soon as the RC is divided into two non-overlapping segments, $[0,\alpha]$ and $[\beta,1]$, continued optimization will collapse the segments into 0 and 1 by sending $\alpha \rightarrow 0$ and $\beta  \rightarrow 1$. However, even an approximate RC, obtained just before the RC is divided into two segments can be useful. It is possible, when such a partial optimization of the RC increases the fraction of points with correct values of RC. In this case, a new ensemble of many short trajectories is prepared, by starting them from points selected uniformly along the RC. The new ensemble, will have a higher fraction of points with higher values of $\alpha$ and smaller values of $\beta$. By iterating this process, one can converge to the ensemble with points uniformly sampled along the RC. This process is somewhat analogous to the way uranium is enriched in centrifuges: each cycle leads only to a marginal increase in the concentration of the desired isotope. However, by repeating the cycle many times, the concentration gets exponentially increased. We call this approach the transition state ensemble enrichment, TSEE. 

\begin{figure}[htbp]
	\centering
	\includegraphics[width=.5\linewidth]{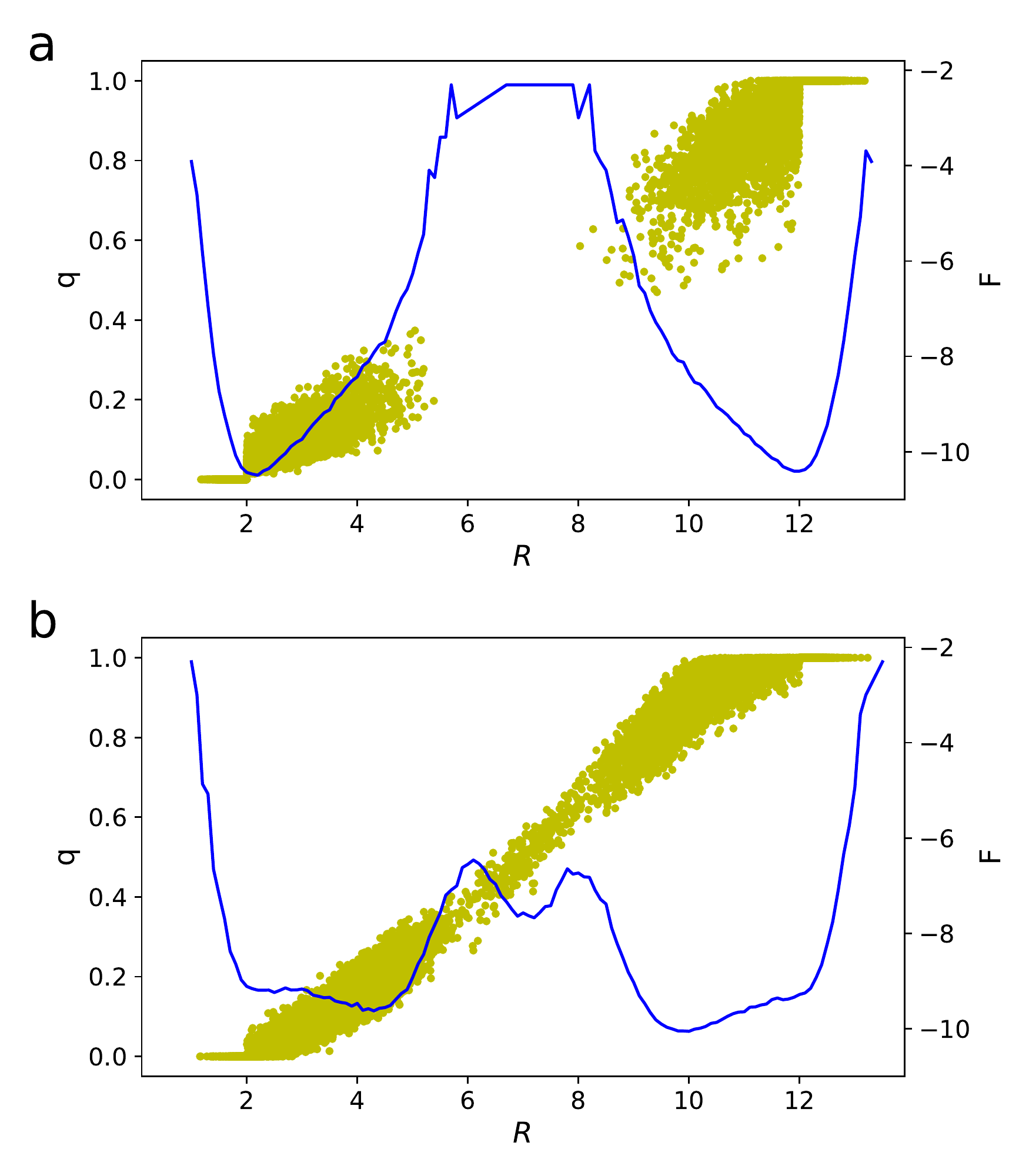}
	\caption{Convergence of the TSEE approach in application to the model system. Initial distribution of points along an OP $F(R)$ (blue line); two-dimensional distributions of points on the q-R plane (yellow dots). Panel a and b show first and second iterations, respectively. For details see text.}
	\label{fig:TSEE}
\end{figure}

We illustrate the TSEE approach on the model 50 dimensional system (Fig. \ref{fig:TSEE}). \textbf{First Iteration}. We start by sampling the boundary states. 10000 short trajectories of length 10 with timestep of $0.1$ are simulated by starting from points with $R=2$ and $R=12$. This is done by assigning the 50 coordinates to random numbers uniformly distributed in the range [-0.5,0.5], and re-scaling them so that $R$ equals to 2 or 12. $F(R)$ on Fig. \ref{fig:TSEE}a shows that the points are distributed mainly around $R=2$ and $R=12$, with almost no points in the TS region. The NPNEq algorithm is applied to optimize the putative RC. The degree of polynomial initially is set at 2 to limit the flexibility of RC to avoid its quick division into two segments. The degree is gradually increased during optimization. NPNEq optimization is continued either until a segment $0.45<q<0.55$ had fewer than 100 points, i.e., the RC is close to be divided into two segments (stopping condition one), or until the iteration number reached 1000 (stopping condition 2). The NPNEq optimization exited after 566 iterations with stopping condition one. The distribution of points on the q-R plane on Fig. \ref{fig:TSEE}a shows that optimization has stratified points according to the putative committor or that the initial and final parts of the committor are determined relatively accurate. By selecting points from the different regions along $q$, different regions of the configurations space can be sampled more uniformly.

\textbf{Second Iteration}. 10000 points are drawn uniformly along $\tilde{q}$ - the putative committor transformed to the natural coordinate. These points are used to start 10000 short trajectories of length $10$ with time step of $0.1$. $F(R)$ on Fig. \ref{fig:TSEE}b shows that some of these trajectories visited the TSE. The same NPNEq algorithm is applied to optimize the putative RC. The algorithm terminated after 1000 iterations with stopping condition two, i.e., the optimization is robust with no division of the RC. The distribution of points on the q-R plane on Fig. \ref{fig:TSEE}b shows that they cover all of the committor. Thus, the TSEE algorithm converged on the second iteration.

\section{Concluding Discussion}
We have described non-parametric non-equilibrium approaches to accurately determine the committor function and re-weighting factors from non-equilibrium simulations. Given a representative and extensive sampling of the configuration space, e.g., a large ensemble of short trajectories, the proposed approaches can be used to determine the equilibrium free energy profile as a function of the committor. The profile, together with the position dependent diffusion coefficient, specify a diffusive model of the equilibrium dynamics. The model can be used to compute the following important properties of the dynamics \textit{exactly}: the equilibrium flux, the mean first passage times, and the mean transition path times between any two points on the committor \cite{krivov_protein_2018, banushkina_optimal_2016}. The power of the approach was illustrated on a model 50-dimensional system and a realistic protein folding trajectory. 

In application to the eigenvectors optimization problem, the obtained equations are similar to those obtained in the EDMD approach  \cite{wu_vamp_2017, williams_data-driven_2015}. Here, however, these equations describe a single iteration of the iterative optimization process, which leads to the following advantages. A major weakness of the parametric approaches, e.g., those using a linear combinations of molecular descriptors/features or a deep neural network, is the choice-of-basis (choice of functional form) problem. While it was argued that ”the expressive power of neural networks provides a natural solution to the choice-of-basis problem” \cite{hernandez_variational_2018}, finding an optimal architecture of a neural network and input variables are difficult tasks. While intuition can help to solve the problem for low-dimensional model systems, the difficulty in the case of complex realistic systems becomes apparent, when one remembers that such a function should be able to accurately project a few million snapshots of a very high-dimensional trajectory. In particular, it implies an extensive knowledge of the system, and that an acceptable solution is likely to be system specific. The developed non-parametric approaches can approximate any reaction coordinate with high accuracy. While each iteration may depend on the exact choice of the family of collective variables/molecular descriptors/features, the final reaction coordinate does not, since it provides optimum to a (non-parametric) target functional, when the optimization converges. We assume, of course, that the employed input variables provide all the important information. For the analysis of biomolecular simulations one can suggest the inter atom distances, or the sines and cosines of internal angles as the standard sets of input variables.   
The developed non-parametric approaches are able to accurately approximate the committors and eigenvectors of realistic systems at the shortest timescales of trajectory sampling interval of $0.2$ ns \cite{krivov_protein_2018,krivov_blind_2020}. Also, one of the reasons of using re-weighted data in eigenvector approximation \cite{wu_vamp_2017} is to avoid complex eigenvalues/eigenvectors since they lack interpretability. This strategy assumes that the re-weighting factors can be accurately determined for every trajectory point, which is a difficult task. The problem of complex values, however, has a simple solution in the iterative optimization. First, since the number of basis functions used during each iteration is rather small, the statistical noise is small and the occurrence of complex eigenvalues/eigenvectors is an infrequent event. Thus one can either skip such an iteration or accept it, truncating complex variables to real parts. 

Note, that while we call the approaches non-parametric, emphasizing that we focus on RC as a function of trajectory (trajectory time or trajectory snapshot), $r(i \Delta t_0)$, rather than as a function of configuration space, $r(\bm{X})$, it is possible to record all the RC transformations during iterative optimization (training) and apply them later to new (test) data, e.g., for cross-validation. Alternatively, one can perform cross-validation on the fly, by computing parameters of RC transformations on the train part of the data, while applying these transformations to the train and test parts of the data. It can be trivially implemented by setting $I_t(t)=0$ for the test data. 

It is instructive to compare different descriptions of molecular dynamics, e.g., using committors vs using eigenvectors as reaction coordinates for free energy landscapes or using eigenvectors to approximate the evolution (forward, backward or Koopman) operators of the dynamics. They all have strong and week points. For example, if it is sufficient to know just such important quantities of dynamics as the equilibrium flux, the mean first passage times or the mean transition path times between two states of interest, e.g., folded and unfolded states or bound and unbound states, then the diffusive model along the committor allows one to determine these properties exactly (between any two points along the committor). This result is valid for any system, irrespective of complexity of its free energy landscape, and does not assume the separation of timescales \cite{krivov_reaction_2013,krivov_protein_2018}. The diffusive model can be used to determine, rigorously and in a direct manner, the free energy barrier and the pre-exponential factor - the major determinants of molecular kinetics \cite{krivov_protein_2018}. Distribution of transition-path times is an example of quantity that can not be accurately determined from the diffusive models, in general \cite{satija_broad_2020}. If one assumes the separation of timescales, then the projected dynamics becomes Markovian, and the diffusive model provides it complete description. A set of slowest eigenvectors/eigenfunctions (basis) can provide a close approximation to the evolution operators and thus can be used to compute accurately many properties of the dynamics. One, however, may require a relatively large basis set to accurately estimate the quantities, that can be computed exactly by the diffusive model along the committor, that requires the determination of just one optimal coordinate. Also, it is not straightforward to visualize an approximated evolution operator, while a free energy landscape as a function of one or two optimal reaction coordinates, provides a clear, intuitive and quantitative picture of the dynamics. Another difference is that iterative optimization of committors is robust, while that of eigenvectors has an inherent instability \cite{krivov_blind_2020}. In the committor case, one seeks an optimal coordinate between two given states (a variant of supervised learning or rather reinforcement learning). The eigenvector optimization can be considered as a variant of unsupervised learning: one seeks eigenvectors with smallest eigenvalues, which describe the slowest dynamics. However, some of such eigenvectors are not of interest. For example, in protein folding, such an eigenvector could describe a much slower torsion angle isomerization process \cite{banushkina_nonparametric_2015, mcgibbon_identification_2017}. Another, more likely possibility, is due to a limited sampling, especially in the case of many short trajectories. There are many parts of the configuration space that were visited only once, and eigenvectors describing those transitions have small eigenvalues. Thus, starting with an eigenvector of interest, the iterative approach may eventually converge to an eigenvector, with smaller eigenvalue, but of no interest. To determine the committor, one needs to specify two boundary states. Proper definition of such states is a difficult problem. For example, a natural approach of using the rmsd from a structure may lead to inaccuracies and hide complexity of the free energy landscapes \cite{krivov_blind_2020}. The problem is likely to be more
severe for systems with complex free energy landscape, e.g., intrinsically disordered proteins \cite{lohr_abeta_2021}. One general strategy of blind, unbiased analysis of dynamics, that uses strong points of both eigenvectors and committors is as follows. 
First, eigenvectors, even not completely optimized/converged, are used for an exploratory analysis of free energy landscapes, e.g., to locate and define the boundary states \cite{krivov_blind_2020}. This is followed by the determination of the committors between these states and the corresponding equilibrium free energy profiles. 

The described non-parametric approaches have only two assumptions - representative sampling and that the underlying dynamics is Markovian in the configuration space. For example, for atomistic MD simulations, where the dynamics is Newtonian at the integration time-step, the sampling/saving interval $\Delta t_0$ needs to be sufficiently large, so that the dynamics have no memory about the momenta. In principle, shorter sampling intervals can be employed if dynamics in the phase space is considered, i.e., the committor is a function of positions and momenta, however it is not yet clear, if it can bring significant advantages. Since representative sampling does not need to cover exhaustively the entire configuration space, the approaches do not suffer from the curse of dimensionality. It suggests that these approaches can be used to investigate dynamics of large biomolecular systems in a rigorous and accurate way. The approaches allows straightforward parallelization and can be adapted for exascale computing.

By a representative sampling we mean such a sampling which provides a representative, but not exhaustive/complete sampling of all the important regions of the configuration space.  By an extensive sampling we mean a sampling of such a size that overfitting by the non-parametric approaches is not possible or negligible. In case, when the sampling is not extensive, i.e., some regions do get overfitted, it can be straightforwardly rectified by performing many short simulations starting from the configurations in the overfitted regions. 

We have suggested how one can generate such a representative and extensive sampling in a number of realistic practical scenarios, e.g., in tandem with many developed enhanced sampling techniques. We have also described a generic approach, the transition state ensemble enrichment, TSEE, which generates such a representative and extensive sampling in an iterative, self-consistent manner, by "growing" committor segments towards each other starting from the boundary states.

The developed non-parametric approaches determine values of a specific optimal RC (e.g., the committor) for an ensemble of configurations, without using any system specific information. They can be considered analogous to linear algebra routines (e.g. the LAPACK library \cite{laug}), where given, for example, a matrix, one can obtain numerical values of eigenvector components. Here, however, the task is complicated by the fact that the transition probability matrix is not given explicitly; only an ensemble of trajectories is provided. In particular, one can not compute the matrix-vector product, $Av$, the basic operation in iterative linear algebra methods. Also, the configuration space is continuous, meaning that we are dealing with an infinite-dimensional problem, which is somewhat simplified by considering a large representative sample of points instead. In addition, the non-parametric approaches to determine the eigenvectors, require additional efforts to suppress the 'inherent instability' \cite{krivov_blind_2020}. However, the developed approaches show that these problems are solvable, and further development of the framework should deliver rigorous, robust and efficient tools to solve the sampling problem.

\section{Appendix}
\subsection {Alternative derivations of Eq. \ref{q:a2}}
Another way to derive Eq. \ref{q:a2} is by using the Galerkin condition. Consider a variation of the RC, approximating the committor function, that satisfies the boundary conditions: $x(i)+\delta x(i)=x(i)+\tilde{I}_b(i)\sum_j \alpha_j f_j(i)$, where $x(i)$ satisfies the boundary condition $x(A)=0$ and $x(B)=1$, while $\tilde{I}_b(A)=0$, $\tilde{I}_b(B)=0$ and  $\tilde{I}_b(i)=1$ otherwise, and $f_j$ are the basis functions. The error vector $\epsilon (j)$, or the vector of residuals, is defined as 
\begin{equation}
	\label{q:e}
		\sum_{i}[x(i)-x(j)]n(i|j,\Delta t_0)=\epsilon (j),\, \mathrm{for} \, j \ne A, B
\end{equation}
For the committor function $\epsilon=0$. The optimal variation is defined by the Galerkin condition: the error vector is orthogonal to all the basis functions of the variation $\epsilon \perp f_k$
\begin{equation}
	\label{q:e2}
	\sum_j[\sum_{i}[x(i)+\delta x(i)-x(j) -\delta x(j)]n(i|j,\Delta t_0)] \tilde{I}_b(j)f_k(j)=0,
\end{equation}
where we used $\tilde{I}_b$ to extend the summation to all $j$. This system of equations is translated to the RC time-series as follows
\begin{equation}
\sum_t [r(t+\Delta t) +\delta r(t+\Delta t) -r(t) -\delta r(t)] f_k(t)I_t(t)I_b(t)=0, 
\end{equation}
which leads to Eq. \ref{q:a2}.

Yet another way to derive Eq. \ref{q:a2} is to consider the following optimization functional 
\begin{subequations}
	\label{q:alt}
	\begin{align}
		&\min_x \sum_{j\ne A,B} n(j)[x(j)-\sum_i P(i|j,\Delta t_0)x(i)]^2\\
		&x(A)=0,\quad x(B)=1
	\end{align}
\end{subequations}
The functional equals $\sum _j \epsilon^2(j)/n(j)$ and attains its minimum when $\epsilon(j)=0$, which gives the committor equation (Eq. \ref{q}). The functional does not assume the detailed balance. Minimization of such a functional is a standard approach of solving a linear system of equations (for committor) iteratively. This functional, however, can not be expressed in terms of RC time-series $r(i\Delta t_0)$, and thus can not be used for non-parametric optimization. 
Consider now the modified optimization problem
\begin{subequations}
	\label{q:alt2}
	\begin{align}
		&\left.\min_x \right|_{x'=x} \sum_{j\ne A,B} n(j)[x(j)-\sum_i P(i|j,\Delta t_0)x'(i)]^2\\
		&x(A)=0,\quad x(B)=1
	\end{align}
\end{subequations}
While the entire functional can not be expressed in terms of RC time-series, the part that depends on $x$ can be expressed. The other part is not important, as it depends solely on $x'$ and is fixed during optimization. When expressed in terms of RC time-series, it equals $\sum_t r^2(t)I_t(t)-2r'(t+\Delta t)r(t)I_t(t)$, i.e., it is equal to Eq. \ref{q:i} up to the term $r'(t+\Delta t)^2I_t(t)$, which is also held constant and disappears after differentiation. It means that Eq \ref{q:a2} (the NPNEq algorithm) can be also interpreted as iterative solving of the (more conventional) optimization problem of Eqs \ref{q:alt}.

\subsection{$Z_q$ criterion} 
Consider a functions "conjugated" or time-reversed to $Z_q$
\begin{equation}
\frac{\partial Z_q^T(x,\Delta t)}{\partial x}=\sum_{ij} \delta (x-x(i))(x(j)-x(i))n(i|j, \Delta t)
\label{q:qs}
\end{equation}
For the half-sum of the two functions one obtains
\begin{equation}
\frac{\partial}{\partial x} \frac{Z_q(x,\Delta t)+Z_q^T(x,\Delta t)}{2}=\sum_{ij} [\delta (x-x(j))-\delta (x-x(i))][x(i)-x(j)]n(i|j, \Delta t)/2 
\end{equation}
In order to understand the meaning of the half-sum, consider $n(i|j,\Delta t)$ transitions from $j$ to $i$. Then, if $x(i)>x(j)$, one obtains a rectangular pulse from $x(j)$ to $x(i)$ of height $n(i|j,\Delta t)/2 \times |x(i)-x(j)|$. If $x(i)<x(j)$, one obtains a rectangular pulse from $x(i)$ to $x(j)$ of height $n(i|j,\Delta t)/2 \times |x(i)-x(j)|$. But this is exactly the definition of $Z_{C,1}$ \cite{krivov_reaction_2013}. Thus, $$\frac{Z_q(x,\Delta t)+Z_q^T(x,\Delta t)}{2}=Z_{C,1}(x,\Delta t).$$ For equilibrium dynamics, where $n(i|j,\Delta t)=n(j|i, \Delta t)$, one finds  $Z_q=Z_q^T=Z_{C,1}$.

\subsection{Non-parametric determination of eigenvectors from non-equilibrium sampling} 
The re-weighting factors can also be considered as the components of the first right eigenvector (with $\lambda$=1) of a non-equilibrium version of the transfer operator $n(i|j,\Delta t)/n(i)$:

\begin{equation}
	\label{ev:nij}
\sum_j n(i|j,\Delta t)u(j)=\lambda n(i)u(i)
\end{equation}
They are related to the right eigenvectors of the equilibrium transfer operator $T(i|j,\Delta t)=P(i|j,\Delta t)\pi(j)/\pi(i)=n(i|j, \Delta t)w(j)/[n(i)w(i)]$ as
\begin{equation}
	\label{ev:tij}
\sum_j T(i|j,\Delta t)[u(j)/w(j)]=\lambda [u(i)/w(i)],
\end{equation}
i.e., eigenvectors of the transfer operator can be obtained as eigenvectors of Eq. \ref{ev:nij} divided by the re-weighting factors $w(i)$ (the first eigenvector of Eq. \ref{ev:nij}). The eigenvectors of Eq. \ref{ev:nij} can be found as the solution of optimization problem
\begin{subequations}
	\label{u:i}
	\begin{align}
		&\left.\max_x \right|_{x'=x} \sum_{ij} x(i)n(i|j,\Delta t)x'(j)\\
		&\sum_i n(i)x^2(i)=1
	\end{align}
\end{subequations}
which is translated to RC time-series
\begin{subequations}
	\label{u:rc}
	\begin{align}
		&\left.\max_r \right|_{r'=r} \sum_{t} r(t+\Delta t)r'(t)I_t(t)\\
		&\sum_t r^2(t)I_t(t)=1
	\end{align}
\end{subequations}
Taking RC variations as $r(t)=\sum_j \alpha_j f_j(t)$ and following steps analogous those used to derive the NPNEq equations one obtains the following equations (the generalized eigenvalue problem) for the optimal parameters
\begin{subequations}
	\label{v:a}
	\begin{align}
A_{kj}\alpha_j^\star=\lambda B_{kj}\alpha_j^\star \\
A_{kj}=\sum_t f_k(t+\Delta t)f_j(t) I_t(t)\\
B_{kj}=\sum_t f_k(t)f_j(t)I_t(t)
	\end{align}
\end{subequations}
The left eigenvectors of the transition probability matrix 
\begin{equation}
	\label{v:pij}
\sum_j v(j)P(j|i,\Delta t)=\lambda v(i)
\end{equation}
can be found as the solution to optimization problem 
\begin{subequations}
	\label{v:i}
	\begin{align}
		&\left.\max_x \right|_{x'=x}\sum_{ij} x'(j)n(j|i,\Delta t)x(i)\\
		&\sum_i n(i)x^2(i)=1
	\end{align}
\end{subequations}
which is translated to RC time-series 
\begin{subequations}
	\label{v:rc}
	\begin{align}
		&\left.\max_r \right|_{r'=r} \sum_{t} r'(t+\Delta t)r(t)I_t(t)\\
		&\sum_t r^2(t)I_t(t)=1
	\end{align}
\end{subequations}
with the following equations on optimal parameters
\begin{equation}
	\label{v:at}
	[A^T]_{kj}\alpha_j^\star=\lambda B_{kj}\alpha_j^\star,
\end{equation}
where matrices $\mathbf{A}$ and $\mathbf{B}$ are defined in Eq. \ref{v:a}. If the dynamics is inherently reversible or equilibrium (though the sampling may be not), i.e., $P(i|j,\Delta t)\pi(j)=P(j|i, \Delta t) \pi(i)$, then $T(i|j,\Delta t)=P(j|i, \Delta t)$ and $v(j)$ in Eq. \ref{v:pij} are the right eigenvectors of the transfer operator. Thus, Eq. \ref{v:at} can be used to determine the eigenvectors of the transfer operator without using the re-weighting factors, for the dynamics which is inherently reversible or equilibrium, which is usually assumed for molecular simulations. Eqs. \ref{v:a} and \ref{v:at} are similar to equations for obtaining linear combinations of feature variables best approximating eigenvectors of the Koopman operator \cite{wu_vamp_2017,williams_data-driven_2015}.

Further discussion on how to select basis functions or how to suppress possible instability during iterative optimization of eigenvectors can be found in Ref.~\citenum{krivov_blind_2020}. 

\section*{Supporting Information.} Jupyter notebooks containing the analyses are provided in a single zip archive.


\providecommand{\latin}[1]{#1}
\makeatletter
\providecommand{\doi}
{\begingroup\let\do\@makeother\dospecials
	\catcode`\{=1 \catcode`\}=2 \doi@aux}
\providecommand{\doi@aux}[1]{\endgroup\texttt{#1}}
\makeatother
\providecommand*\mcitethebibliography{\thebibliography}
\csname @ifundefined\endcsname{endmcitethebibliography}
{\let\endmcitethebibliography\endthebibliography}{}

\end{document}